\documentclass[aps,superscriptaddress,nofootinbib,eqsecnum,prd,notitlepage,twocolumn]{revtex4-1} 

\pdfoutput=1

\usepackage{amsfonts}
\usepackage{amsmath}
\usepackage{amssymb}
\usepackage{graphicx,color}
\usepackage{float}
\usepackage{hyperref}
\usepackage{subfigure}
\usepackage{dcolumn}% Align table columns on decimal point
\usepackage{soul}
\usepackage{ulem}

\newcommand\be{\begin{equation}}
\newcommand\ba{\begin{eqnarray}}
\newcommand\ee{\end{equation}}
\newcommand\ea{\end{eqnarray}}

\begin{document}

\title{Minimal Preheating}

\author{Robert Brandenberger}
\email{rhb@physics.mcgill.ca}
\affiliation{Department of Physics, McGill University, Montr\'{e}al,
  QC, H3A 2T8, Canada}

\author{Vahid Kamali}
\email{vkamali@ipm.ir}
\affiliation{Department of Physics, McGill University, Montr\'{e}al,
  QC, H3A 2T8, Canada}
\affiliation{
  Department of Physics, Bu-Ali Sina (Avicenna) University, Hamedan 65178,
  016016, Iran}
\affiliation{
  School of Physics, Insitute for Research in Fundamental Sciences (IPM),
  19538-33511, Tehran, Iran}
  
\author{Rudnei O. Ramos} 
\email{rudnei@uerj.br}
\affiliation{Departamento de Fisica Teorica, Universidade do Estado do
  Rio de Janeiro, 20550-013 Rio de Janeiro, RJ, Brazil }
\affiliation{Department of Physics, McGill University, Montr\'{e}al,
  QC, H3A 2T8, Canada}
  
%%%%%%%%%%%%%%%%%%%%%%%%%%%%%%%%%%%%%%%%%%

\begin{abstract}

An oscillating inflaton field induces small  amplitude oscillations of
the Hubble parameter at the end of inflation.  These Hubble parameter
induced oscillations, in turn, trigger parametric particle production of
all light fields,  even if they are not directly coupled to the
inflaton.  We here study the induced particle  production for a light
scalar field (e.g. the Standard Model Higgs field) after inflation as
a consequence of this effect. Our analysis yields a model-independent
lower bound on the efficiency of energy transfer from the inflaton
condensate to particle excitations.  

\end{abstract}

\maketitle

%%%%%%%%%%%%%%%%%%%%%%%%%%%%%%%%%%%%%%%%%%%%
\section{Induced gravitational preheating effect} 
 
A major challenge for inflationary cosmology \cite{Guth} is to provide
a mechanism to produce a hot gas of particles after the period of
inflation, a period during which the number density of particles
existing before inflation is exponentially diluted. The first attempts
to describe this {\it reheating} process were perturbative
\cite{pert}.  But it was soon realized \cite{TB, DK,KLS1, STB, KLS2}
that the oscillating inflaton condensate (which results after
inflation has ended) can induce parametric resonance excitations of
any field which couples to the inflaton. The analysis of the
efficiency of this process is, however, very model independent (see
e.g. \cite{RHrevs} for reviews of reheating).

Here we discuss a model-independent mechanism which leads to the
production of particles, even if they are not non-gravitationally
coupled to the inflaton field.  The source of this phenomenon is the
oscillating contribution to the pressure which the oscillating
inflaton condensate induces. This oscillating pressure leads to a
oscillating contribution to the mass of any field which has the usual
coupling to gravity, in particular the Standard Model Higgs field.
Our process thus provides a channel to directly produce Standard Model
particles after inflation, and it provides a model-independent lower
bound on the efficiency of reheating.

The gravitational particle production channel which we are discussing
leads to the production of quanta of all scalar fields whose mass is
substantially smaller than the inflaton mass.  Gauge fields will be
excited by the same mechanism, as will fermions, except that in the
case of fermions the particle production will be smaller due to Pauli
blocking.  Since dark matter and Standard Model matter are excited by
the same process, there is a possible connection to the dark matter
coincidence problem, namely the puzzle of why the energy densities of
dark matter and visible matter are of similar magnitude today.

The gravitational reheating mechanism which we employ here was first
studied in the context of dark matter production after inflation in
various works \cite{various}.  In \cite{Mesbah}, the mechanism was
studied in the context of moduli and graviton production during a
early universe phase in which a modulus field is coherently
oscillating. Recently, we \cite{we} studied the decay on an axion
condensate by the same mechanism \footnote{See also \cite{Natalia} for
some early work, and \cite{Chunshan} for an application of a parametric resonance instability to describe the generation of photons from a gravitational wave.}.  Here, we emphasize two new aspects. {}First, we
point out that gravitational preheating leads to a direct and
model-independent production channel for Standard Model
particles. Second, we derive a lower bound on the efficiency of the
energy transfer from the inflaton condensate to particle quanta, and
determine a lower bound on the {\it effective temperature} at the end
of the preheating phase.  We also point out a possible connection to
the {\it dark matter coincidence problem}, the mystery of why the
density of dark matter is comparable to the density of Standard Model
matter.

In the following, we will use natural units in which the speed of
light, Planck's constant and Boltzmann's constant are set to one.  We
work in the context of a spatially flat homogeneous and isotropic
background metric with scale factor $a(t)$, where $t$ is physical
time. The comoving spatial coordinates are $x_i, i = 1,2,3$. The
Planck mass is denoted by $m_{\rm Pl}$.  The Hubble expansion rate is
$H(t) = {\dot{a}(t)}/a(t)$, where an overdot denotes the derivative
with respect to time.  $\rho$ and $p$ stand for energy density and
pressure, respectively.

%%%%%%%%%%%%%%%%%%%%%%%%%%%%%%%%%%%%%%%%%%%%%%%%
\section{Equations of Motion}

In the following we will denote the oscillating scalar inflaton field
by $\phi$, and the scalar field whose particle production we study by
$\chi$. Specifically, $\chi$ can be the Higgs field of the Standard
Model of particle physics. For simplicity, we will neglect
nonlinearities in the scalar field sector, and work in the context of
Einstein gravity. Thus, the Lagrangian ${\cal{L}}_m$ of the matter
sector is
\be 
{\cal{L}}_m  =  \int d^4x \sqrt{-g}  \left[\frac{1}{2}
  \partial_{\mu} \phi \partial^{\mu} \phi +  \frac{1}{2}
  \partial_{\mu} \chi \partial^{\mu} \chi  - V(\phi, \chi) \right], 
\ee
with the potential
\be 
V(\phi,\chi) \, = \, \frac{m_\phi^2}{2} \phi^2 +
\frac{m_\chi^2}{2} \chi^2 \, .
\label{pot}
\ee
In the above, $g$ is the determinant of the metric. Note that inflaton
self couplings or couplings between the inflaton and matter would open
the usual preheating channels and make the overall preheating process
more efficient. We are here aiming for a lower bound on the efficiency
of preheating.

At the end of inflation, all the energy density is stored in the
inflaton field condensate which is beginning to oscillate.  It is useful
to recall the general dynamics of an oscillating scalar field after
inflation.  An oscillating scalar field condensate leads to pressure
oscillating about $p = 0$. When the frequency $\omega$ of the
oscillating field is much larger than the Hubble rate,  
\be 
\omega \, \sim \, \dot\phi/\phi \, \gg \, H \,  , 
\ee
then one can show that \cite{Turner:1983he} the energy density in the
scalar field condensate scales as  
\be 
\rho_\phi \,  \propto a(t)^{-3 \gamma}, 
\ee
where
\be 
\gamma \, = \, 2n/(n+2), 
\ee
for a potential 
\be 
V \, \sim \, \phi^n \, .  
\ee
This leads to the time dependence
\be 
a(t) \, \propto \, t^{2/(3\gamma)} 
\ee
of the scale factor.  Then, the inflaton will have oscillations with a
decreasing amplitude 
\be 
A(t) \, \propto \, a(t)^{-6/(n+2)} \propto t^{-2/n} \, .   
\ee
In the case $n =2$ we can understand this result very easily: the time
average of the pressure of the oscillating scalar field is $p =0$ and
we get a matter-dominated phase of cosmological expansion.

However, in the case of large field inflation models (models in which
the slow-roll trajectory is a local attractor in initial condition
space - see \cite{RHB-IC-Rev} for a review), the intial amplitude of
oscillation is ${\cal{A}} \sim m_{\rm Pl}$ and hence the intial value
of $H$ is of the same order of magnitude as the oscillation frequency
$\omega = m_{\phi}$. In this case, the averaging of the
energy-momentum tensor of the inflaton field is not justified, and we
obtain instead the oscillatory corrections which lead to the resonance
effects which we describe below.

Let us then consider the evolution equation for the momentum modes of
a spectator scalar field,  $\chi_k(t)$,  in the background of the
oscillating inflaton field in an expanding universe. The equation of
motion for $\chi_k$ is given by
\begin{equation} 
{\ddot{\chi_k}} + 3 H {\dot{\chi_k}} + \left(\frac{k^2}{a^2} +
m_\chi^2\right) \chi_k  \, = \, 0 \, .  
\end{equation}
It is convenient to consider the rescaled field modes $\psi_k$ defined
via 
\be 
\psi_k \, \equiv  \, a^{3/2} \chi_k \, , 
\label{rescaling}
\ee
whose equation of motion is 
\begin{equation} 
{\ddot{\psi_k}} +  \omega_k^2(t) \psi_k \, = \, 0,  
\label{psieq}
\end{equation}
where
\begin{equation}
 \omega_k^2(t) \, = \, \frac{k^2}{a^2} + m_\chi^2 - \frac{3}{4}
 \left(\frac{{\dot{a}}}{a}\right)^2 - \frac{3}{2} \frac{{\ddot{a}}}{a}
 \, .
\label{omegakt}
\end{equation}
Note that the mass term of the field in Eq.~(\ref{omegakt}) is
modified by the expansion.

The contribution to the mass of $\chi$ due to the expansion in
Eq.~(\ref{omegakt}) can be expressed as
\begin{eqnarray}
- \frac{3}{4} \left(\frac{{\dot{a}}}{a}\right)^2 - \frac{3}{2}
\frac{{\ddot{a}}}{a} \, = \, -\frac{9}{4} H^2 + \frac{3}{2} \dot H^2
\, = \, 6 \pi G p_\phi,
\label{term}
\end{eqnarray}
where we have used that 
\be 
\dot H \, = \, - 4 \pi G (\rho_\phi+p_\phi),
\ee
and where
\be 
p_\phi \, \equiv \, \frac{1}{2} \dot \phi^2 - \frac{1}{2} m_\phi^2
\phi^2 
\ee
is the pressure density of the inflaton field.  The oscillations of
the inflaton field lead to  important effects for the $\chi$-field
mode evolution in Eq.~(\ref{psieq}).

The oscillating homogeneous inflaton field can be parameterized as
\be 
\phi(t) \, = \, A(t) \cos(m_\phi t) \, ,  
\ee
with a decreasing time dependent amplitude $A(t)$. The effects of the
oscillating pressure on the scale factor are of the order $G
A^2(t)$. For small values of the field amplitude, they would be highly
Planck suppressed. However,  the initial amplitude of the field
oscillation at the end of inflation is of the order of the Planck
mass, and hence the effects can be important. 

We perform a perturbative analysis in $G {\cal{A}}^2$, where
${\cal{A}}$ is the initial amplitude of the field oscillation.   As is
clear from (\ref{psieq}) and (\ref{term}), the effects of the
oscillating pressure on the evolution of $\psi_k$ are of the order of
$G {\cal{A}}^2$, and hence, to leading order in the expansion
parameter, we can consider the amplitude of the field oscillation to
decay like in a matter-dominated universe, i.e. given  by 
\be 
A(t) \, = \, {\cal{A}} \,\frac{t_R}{t} \, ,  
\ee
where $t_R$ denotes the time when the oscillations start. This leads
to the following expression for the pressure
\ba 
p_{\phi} \, &=& \, - \frac{1}{2} m_{\phi}^2 A^2
    {\rm{cos}}(2m_{\phi}t)  - A^2 \frac{m_{\phi}}{t}
    {\rm{sin}}(2m_{\phi}t) 
\nonumber \\ 
& & + \frac{A^2}{t^2} {\rm{cos}}^2(m_{\phi}t) \, .  
\ea
In the above, the first term dominates for $m_{\phi} t > 1$, i.e. when
the mass of the inflaton field is larger than the Hubble scale at the
beginning of the period of oscillation.
  
Hence, the $\chi_k$ field modes acquire a time dependent mass and the
time dependence frequency for the field modes becomes
\ba 
\omega_k^2(t)  &=&  k^2 \left( \frac{t_R}{t} \right)^{4/3} +
m_\chi^2 \nonumber \\ 
&-&  3 \pi G {\cal{A}}^2 {\rm{cos}} (2 m_\phi t)
m_{\phi}^2 \left( \frac{t_R}{t} \right)^2 \, .  
\label{omegak}
\ea
Note that for half of the oscillation period,  the oscillating
pressure leads to a tachyonic contribution to the mass term. This term
will dominate for infrared modes with
\be 
\label{IRcond} k  < \sqrt{3 \pi G {\cal{A}}^2} \, \left(
\frac{t_R}{t}\right)^{1/3} m_{\phi} \, , 
\ee
and provided that the mass of the $\chi$ particles is small
\be 
m_{\chi}  <   \sqrt{3 \pi G {\cal{A}}^2} \, \frac{t_R}{t} m_{\phi}\, , 
\ee
which is trivially satisfied for the Higgs field since the Higgs mass
is so many orders of magnitude smaller than the inflaton mass. If
these conditions are satisfied, we expect rapid $\chi$ particle
production which will lead to efficient reheating. We denote the
limiting value of $k$ where (\ref{IRcond}) is satisfied by $k_{max}$.
Note that $k_{max}$ is a decreasing function of time.

%%%%%%%%%%%%%%%%%%%%%%%%%%%%%%%%%%%%%%%%%%%%%%%%
\section{A Lower Bound on the Efficiency of Reheating}
 
Let us consider a low mass field (e.g. the Standard Model Higgs
field), and infrared modes. In this case the equation of motion
(\ref{psieq}) becomes (introducing the rescaled time variable $z = 2
m_{\phi} t$)
\be 
\label{psieq2} \psi_k^{\prime \prime} -  \frac{3 \pi}{4} G
    {\cal{A}}^2  \frac{z_R^2}{z^2}  {\rm{cos}}(z) \psi_k  =  0,
\ee
where a prime denotes the derivative with respect to $z$, and $z_R$ is
the value of $z$ at the time $t_R$. This is the key equation in our
analysis. We call the second term in the above equation the {\it
  driving term}.
 
If it were not for the $z^{-2}$ factor,  Equation (\ref{psieq2}) would
be of the form of a Mathieu equation \cite{Mathieu} and lead to broad
band parametric resonance (i.e. all infrared modes obeying
(\ref{IRcond}) would be exponentially excited. In the following
section we will investigate the solutions of this equation
numerically. Here, we want to provide an analytical lower bound on
particle production.  We do this by making two approximations. In each
step,  we simplify the form of the driving term by using a lower bound
for it.

Notice that the driving term is positive for half the period and
negative for the other half. When it is positive, $\psi_k$ will
experience a tachyonic instability, when it is negative it will
oscillate. Hence, to obtain the growth in the amplitude of $\psi_k$ we
will set the driving term to zero except in the time range $-\pi/4 < z
< \pi/4$, and in this range we use the lower bound $1/2$ on the
amplitude of the oscillatory term. The second approximation is to
neglect the growth of $\psi_k$ during the rest of the oscillation
period of the condensate.

Based on our first approximation, and setting $t = t_R$, in the
z-interval
\be 
\Delta z \, = \, \frac{\pi}{2},
\ee
the equation (\ref{psieq2}) can be approximated by
\be 
\psi^{\prime \prime} \, = \,  \frac{3}{8} \pi G {\cal{A}}^2
\psi_k \, .  
\ee
The solution is exponentially increasing with Floquet exponent $\mu_k$
given by
\be 
\mu_k^2 \, = \, \frac{3}{8} \pi G {\cal{A}}^2 \, , 
\ee
and thus the amplitude of $\psi_k$ increases during one period by
\be 
\psi_k(z_i + 2\pi) \, = \, \psi_k(z_i) e^{\pi \mu_k / 2 } \, , 
\ee
where $z_i$ is the value of $z$ at the beginning of the period.  Thus,
the effective growth rate is $1/4$ of the instantaneous growth rate
given by $\mu_k$.

To determine the efficiency of the resonance, we need to compare the
growth rate ${\tilde{\mu}}_k$ with the Hubble expansion rate $H$,
where ${\tilde{\mu}}$ is the growth rate in terms of physical time $t$
\be 
{\tilde{\mu}}_k \, = \, 2 m_{\phi} \mu_k \, .  
\ee
Recalling that the effective growth rate ${\tilde{\mu}}_{k, eff}$ is
$1/4$ of the maximal growth rate given by $\mu_k$, and making use of
the Friedmann equation to express $H$ in terms of the inflaton
condensate energy density we find that
\be 
{\tilde{\mu}}_{k, eff} \, = \, \frac{3}{8 \sqrt{2}} H \, .  
\ee
Note that both sides scale in the same way as a function of time. Thus,
we see that the resonance is not strong.  The growth of $\psi$ can
therefore be approximated as a linear growth in time
\be 
\psi_k(t) \, \simeq \, (t - t_R) {\tilde{\mu}}_{k, eff}
\psi_k(t_R) \, .  
\ee
In terms of the canonical field $\chi$ this means (using the scaling
of $a(t)$ in a matter dominated epoch)
\be 
\label{chimode} \chi_k(t) \, \simeq \, t_R {\tilde{\mu}}_{k, eff}
\chi_k(t_R),
\ee
for $t \gg t_R$.  In the next section we will see that the numerical
analysis confirms the linear growth in time of $\psi_k(t)$.

Before discussing the numerical analysis of the equation of motion
(\ref{psieq}) with frequency given by (\ref{omegak}) we provide an
estimate on the efficiency of the energy transfer. We assume that all
$\psi_k$ modes begin in their quantum vacuum state with amplitude
\be 
\psi_k(t_R) \, = \, \frac{1}{\sqrt{2k}} \, , 
\ee
which is a natural assumption since inflation redshifts all field
excitations and leaves behind a matter vacuum state.

A lower bound on the energy density of $\chi$ particles at time $t$
can then be obtained by integrating the contribution of all Fourier
modes of $\chi$ up to the time dependent cutoff $k_{max}(t)$ from
(\ref{IRcond})
\be 
\label{chienergy} \rho_{\chi}(t) \, \sim \, 4 \pi
\frac{1}{(2\pi)^3} \int_{k = 0}^{k_{max}(t_R)} dk k^4
\frac{a^2(t_R)}{a^2(t)} \chi_k(t)^2 \, , 
\ee
where two of the factors of $k$ are from the phase space of modes, and
the other two (plus the scale factor term) from the gradient
contribution to the mode energy.  Combining (\ref{chimode}) and
(\ref{chienergy}, and inserting the values of $k_{max}(t)$ and
${\tilde{\mu}}_{k, eff}$ we obtain
\be 
\label{chienergy2} \rho_{\chi}(t) \, \sim \, \frac{27}{512} \pi (G
    {\cal{A}}^2)^3 \left(\frac{t_R}{t} \right)^{4/3}  m_{\phi}^4 (
    m_{\phi} t_R)^2 \, .  
\ee
The upper bound on the time $t$ when the energy transfer from the
inflaton condensate to $\chi$ particle quanta is complete can be
obtained by setting the result (\ref{chienergy2}) equal to the energy
density 
\be 
\rho_{\phi}(t) \, = \, \frac{1}{2} m_{\phi}^2 {\cal{A}}^2 \left(
\frac{t_R}{t} \right)^2  ,
\ee
of the inflaton condensate density.  

The {}Friedmann equation yields the relation
\be 
m_{\phi}^2 t_R^2 \, = \, \frac{1}{3 \pi} \left( G {\cal{A}}^2
\right)^{-1} \, , 
\ee
and inserting this result into (\ref{chienergy2}) yields the
expression
\be 
\label{endtime} \left( \frac{t}{t_R} \right)^{2/3} \, \sim
\frac{266}{3} \left( G {\cal{A}}^2 \right)^{-1} \left( G m_{\phi}^2
\right)^{-1} ,  
\ee
for the time when the energy transfer is complete.
  
We can define an ``equivalent temperature'' \footnote{Note that the
quanta produced do not have a thermal distribution. This is always the
case at the end of a period of preheating.} $T_{RH}$ via
\be 
T_{RH}^4 \, = \, \rho_{\chi}(t) \, .  
\ee
and the time $t$ given by (\ref{endtime}). Equivalently,
\be 
T_{RH}^4 \, = \, \rho_{\phi}(t) \, = \, \frac{1}{2} m_{\phi}^2
    {\cal{A}}^2  \left( \frac{t_R}{t}  \right)^2 \, .  
\ee
Hence, from (\ref{endtime}) we find
\be \label{Result} T_{RH} \, = \, \left( \frac{1}{\sqrt{2}} m_{\phi}
    {\cal{A}} \right)^{1/2} \left( \frac{3}{266} \right)^{3/4}  \left(
    G {\cal{A}}^2 \right)^{3/4} \left( G m_{\phi}^2 \right)^{3/4} \, .
\ee
The first term on the right hand side of (\ref{Result}) is the maximal
temperature $T_{max}$ after inflation (obtained for instantaneous
transfer of the inflaton energy into particles). The other three
factors yield the suppression of the actual reheating temperature
compared to this maximal value.  For $G {\cal{A}} \sim 1$ and for the
characteristic value $G m_{\phi} \sim 10^{-6}$ for standard single
field inflation models (corresponding to an inflationary energy scale
of about $10^{16} {\rm{GeV}}$ we find the lower bound
\be 
\label{eval} T_{RH} \,  \sim \, 10^{-6} T_{max} \, .  
\ee
Equations~(\ref{Result}) and (\ref{eval}) are the main results of our
work.

%%%%%%%%%%%%%%%%%%%%%%%%%%%%%%%%%%%%%%%%%%%%%%%%
\section{Numerical Analysis}
 
Since the time scale of our resonance phenomenon is of the same order
of magnitude as the Hubble expansion time scale, it is crucial to
confirm our analytical estimates by numerically solving the key
equation (\ref{psieq}) with frequency given by (\ref{omegak}).  With
the numerics we can study the evolution for general values of $k$ and
$m_{\chi}$.

\begin{center}
\begin{figure}[!htb]
\includegraphics[width=6.4cm]{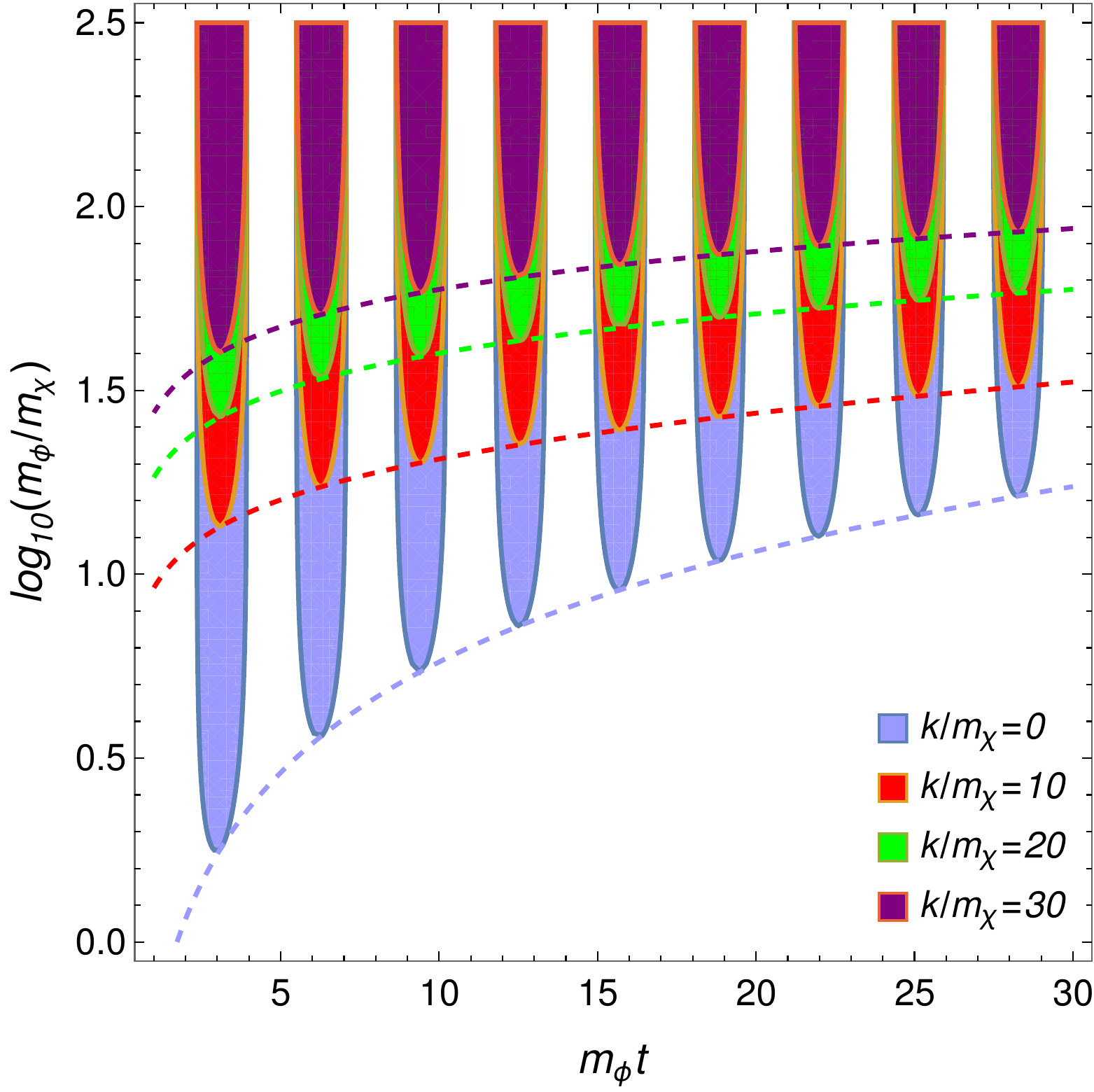}
\caption{Parameter values and intervals of time for which
  $\omega_k^2<0$, indicated by the colored regions.  The vertical axis
  is the ratio of the inflaton mass to the mass of the $\chi$ field.
  Roughly speaking, only for $\chi$ masses smaller than the $\phi$
  mass can there be an instability. The different colors show the
  results for different values of $k$.  For values of $k$ larger than
  $m_{\chi}$, it is the value of $k$ which determines if an
  instability is possible. It is infrared modes which can undergo the
  resonant excitation.  The dashed lines indicate the lowest value of
  $m_{\phi} / m_{\chi}$ for which there can be a resonance.}
  \label{fig1}
\end{figure}
\end{center}

We can choose, without loss of generality,  the initial time $t_R$ as
the time the inflaton field starts  oscillating after inflation.  For
slow roll inflation models, this is given by
\be 3 H(t_R) \, \equiv \, \frac{2}{t_R} \, = \, m_\phi \, ,
\label{tR}
\ee
which yields $t_R=2/m_\phi$. From Eq.~(\ref{omegak}), we can see that
for small values of $k$ and $m_{\chi}$ (relative to $m_\phi$) there
will be time intervals when $\omega_k^2$  becomes negative.  Whenever
this happens, the modes $\psi_k$ grow exponentially.  This is
illustrated in {}Fig.~\ref{fig1} which shows the parameters and times
for which $\omega_k^2<0$ and where this exponential growth of $\psi_k$
can happen.

%%%%%%%%%%%%%%%%%FIGURE01%%%%%%%%%%%%%%%%%%%
%%%%%%%%%%%%%%%%%%%%%%%%%%%%%%%%%%%%%%%%%%%%

{}From {}Fig.~\ref{fig1} we can see that for any sufficiently small
values of $k$ and sufficiently small ratios for the masses,
$m_\chi/m_\phi \ll 1$, the square of the frequency for the $\chi_k$
modes becomes negative. This happens twice every complete cycle of
oscillation,  in a time interval $ -\pi/4 < m_\phi \Delta t < \pi/4$.
Since the maximal value of $k$ for which we expect resonance to
happen, namely the $k_{max}$ given in (\ref{IRcond}), is a decreasing
function of time,  there willl be, for any value of $k$,  a maximum
time $t_{\rm max}$ for which $\omega_k^2$ can become negative.  This
maximal time (which depends on both $m_\phi$ and $m_\chi$) is
indicated by the dashed lines in {}Fig.~\ref{fig1}.  {}For $t > t_{\rm
  max}$  there is no longer any exponential growth of $\chi_k$, and
the modes simply oscillate after that time with the amplitude reached
after the last instability region is crossed.  This maximum time
$t_{\rm max}$ is given by setting the cosine in Eq.~(\ref{omegak}) as
one (i.e., at its maximum value) and solving the resulting equation
for $\omega_k(t_{\rm max}) = 0$.  This yields
\ba 
& & m_\phi t_{\rm max} \, = \,  \nonumber \\ & & \left[3 \frac{m_
    {\phi }^2}{m_ {\chi }^2}+\frac{2 k^2 F(k,m_\phi,m_\chi)}{m_\chi^2
    \left(\frac{m_\chi}{m_\phi}\right)^{2/3}}- \frac{8 k^4
    \left(\frac{m_\chi}{m_\phi}\right)^{2/3}}{3 F(k,m_\phi,m_\chi)
    m_\chi^4} \right]^{1/2} \, , \nonumber\\
\label{tmax}
\ea
where $F(k,m_\phi,m_\chi)$ in the above equation is given by
\be 
F(k,m_\phi,m_\chi) \, = \, 3^{1/3}\left(-1 + \sqrt{ 1+ \frac{64
    k^6}{243 m_\phi^4 m_\chi^2}}\, \right)^{1/3}.
\label{functionF}
\ee
Note that for small values of $m_{\chi}$ (compared to $m_{\phi}$) the
back-reaction of particle production on the inflaton condensate will
shut off the resonance before $t_{max}$. This back-reaction time scale
was discussed at the end of the previous section.

Neglecting the abovementioned back-reaction effect, the number of
times an instability region is crossed for given values of $k,\,
m_\phi$ and $m_\chi$ is given  by 
\be 
\Delta N \, = \, m_\phi t_{\rm max}/\pi \, .  
\ee
{}For $k\ll m_\chi$, we have for instance that  
\be 
\Delta N \, \sim \, m_\phi/(\pi m_\chi) \, .  
\ee
The smaller the ratio $m_\chi/m_\phi$, the larger will be the growth
of the momentum modes for the $\chi$ field. The small oscillations of
the metric induced by the oscillating inflaton can, therefore,  as
discussed in the previous section, be an efficient way of producing
$\chi$-particles.  Let us investigate this effect in more detail.  

\begin{center}
\begin{figure}[!htb]
\includegraphics[width=6.4cm]{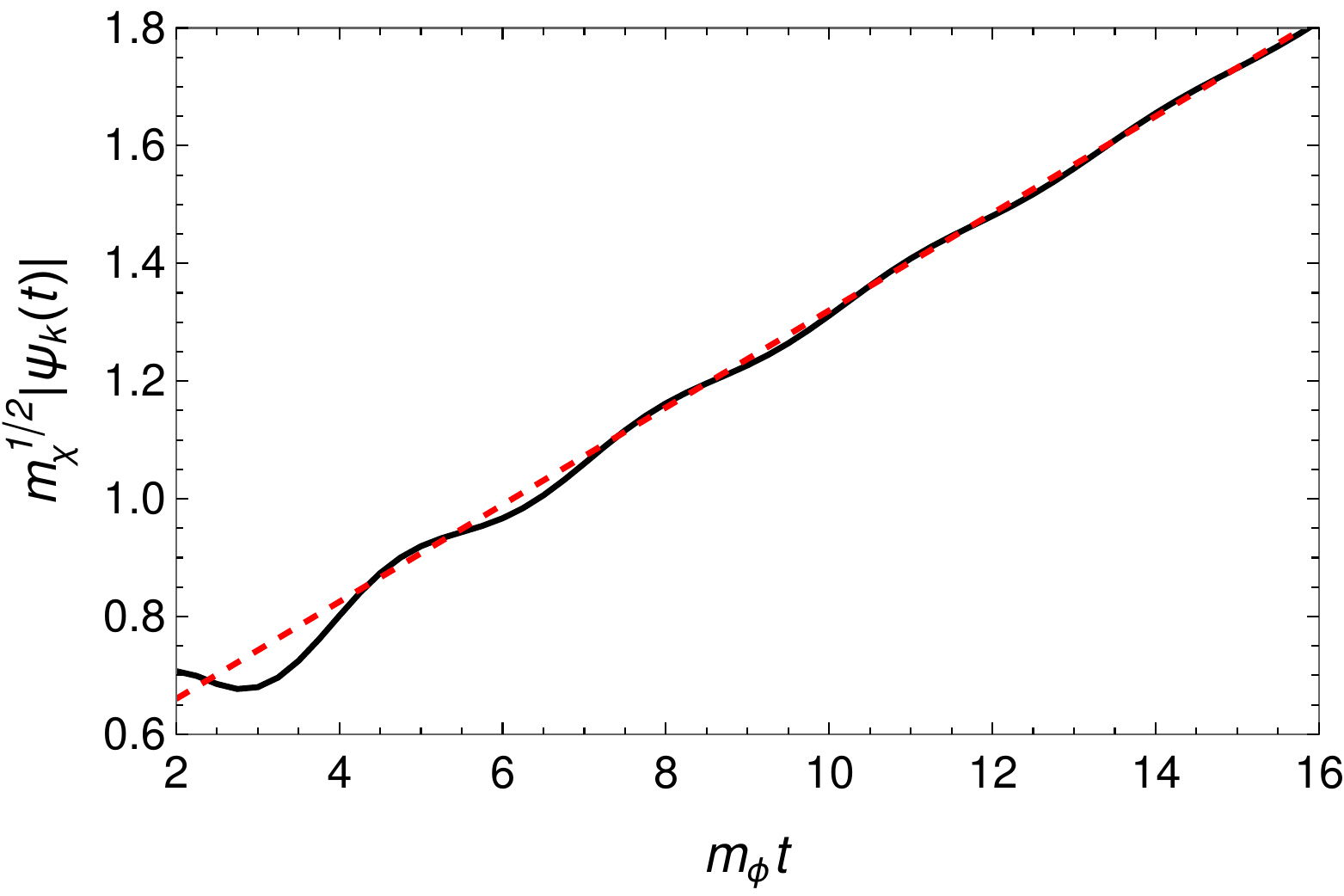}
\caption{Time evolution of $\psi_k$ for $k=0$ and $m_{\chi}=10^{-6}
  m_\phi$ (solid black curve).  The fit to a linear growth (red dashed
  curve) as a function of time is a good one, and it is found to be
  given by $|\psi_k(t)|\sim 0.0825 (m_\phi t +6)/m_\chi^{1/2}$.  We
  have used the usual vacuum initial conditions when solving for the
  mode equations, i.e., $\psi_k(t_0) = e^{i \omega_k(t_0)
    t_0}/\sqrt{2\omega_k(t_0)}$ and $\dot{\psi}_k(t_0) = i e^{i
    \omega_k(t_0) t_0} \sqrt{\omega_k(t_0)/2}$ and with $t_0$ set at
  $t_R=2/m_\phi$ [see Eq.~(\ref{tR})].  }
  \label{fig2}
\end{figure}
\end{center}

In {}Fig.~\ref{fig2} we consider the evolution of $\psi_k$ as a
function of time $t$, in the case $k=0$ and $m_{\chi} \ll m_{\phi}$.
As is evident, there is indeed resonant growth of the amplitude, as
expected from the analytical analysis of the previous section. As
expected based on our analytical approximations, the resonant growth
is linear in time.

%%%%%%%%%%%%%%%%%FIGURE02%%%%%%%%%%%%%%%%%%%
%%%%%%%%%%%%%%%%%%%%%%%%%%%%%%%%%%%%%%%%%%%%

\begin{center}
\begin{figure}[!htb]
\subfigure[]{\label{fig3a} \includegraphics[width=6.4cm]{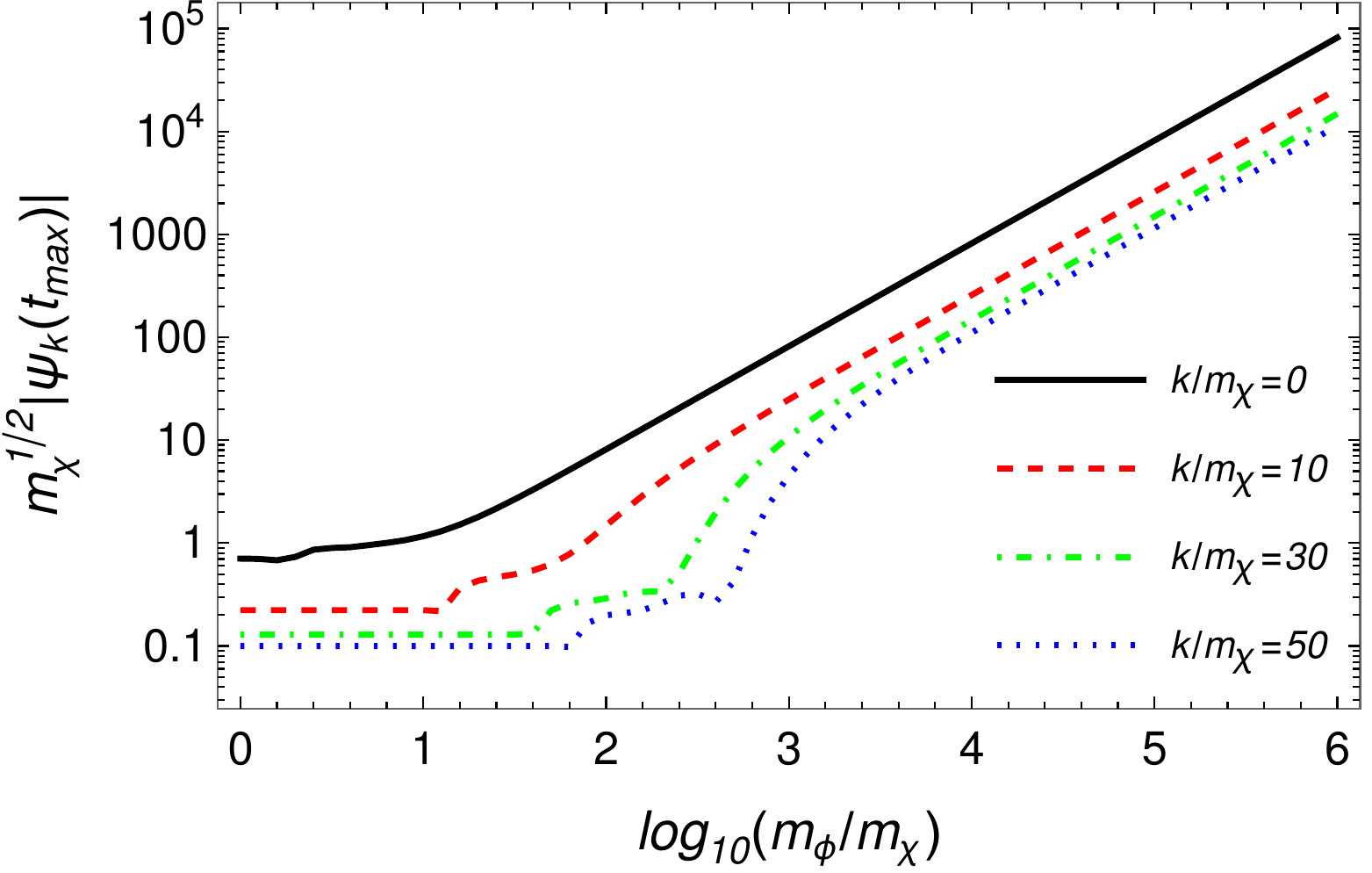}}
\subfigure[]{\label{fig3b} \includegraphics[width=6.4cm]{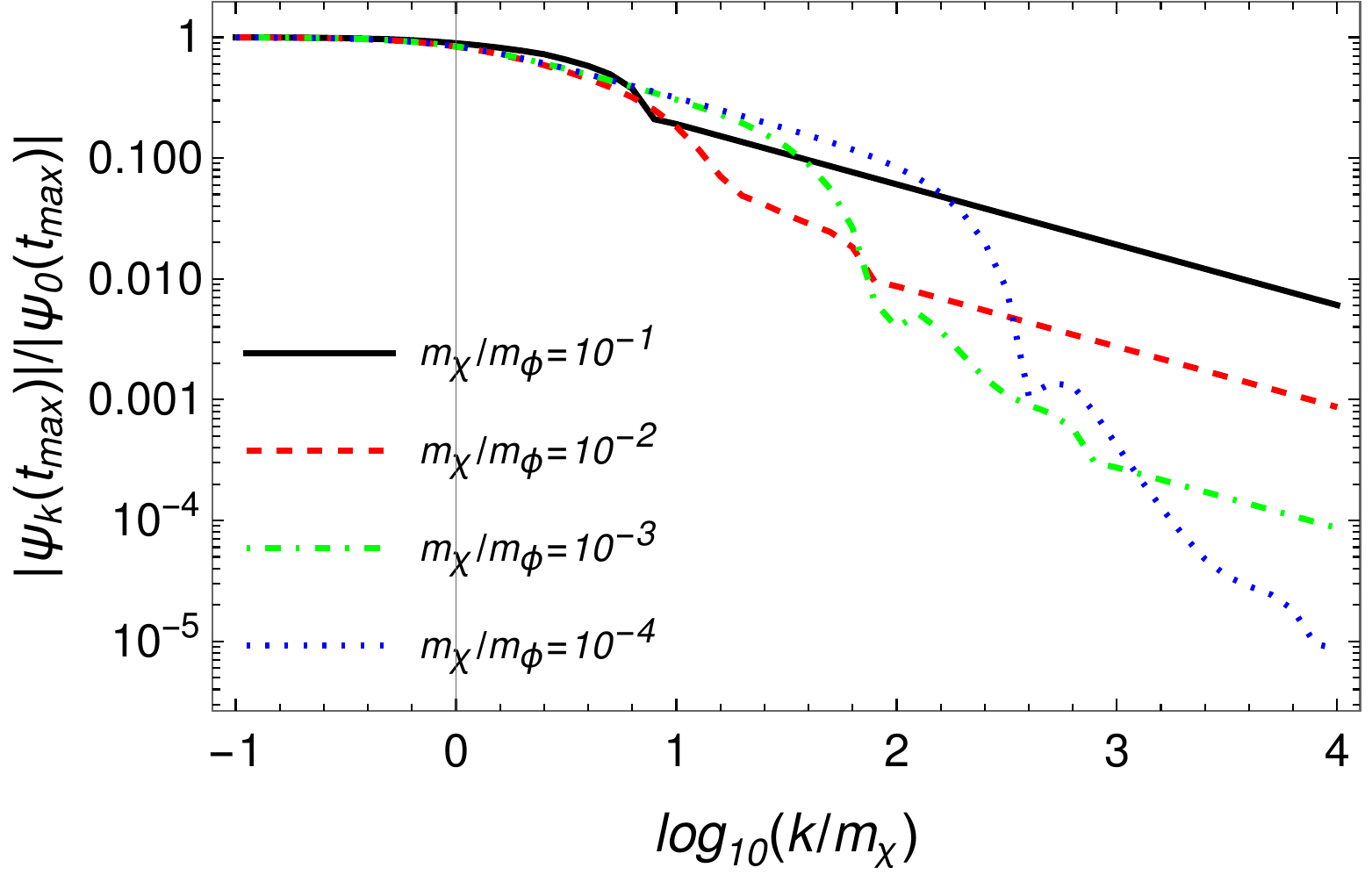}}
\caption{The modulus of $\psi_k(t)$ evaluated at the maximum time
  $t=t_{\rm max}$  and for different values of the ratio of masses
  $m_\phi/m_\chi$ and values of the momentum $k$. The evolution was
  started with vacuum initial conditions as in {}Fig.~\ref{fig2}.}
  \label{fig3}
\end{figure}
\end{center}

In {}Fig.~\ref{fig3}, we show the numerical solution of the
$\chi$-field momentum mode equation, Eq.~(\ref{psieq}), using
Eq.~(\ref{omegak}), with $\psi_k(t)$ evaluated at the maximum time
$t=t_{\rm max}$ at which the maximum amplitude for $\psi_k(t)$ is
reached.

%%%%%%%%%%%%%%%%%FIGURE03%%%%%%%%%%%%%%%%%%%
%%%%%%%%%%%%%%%%%%%%%%%%%%%%%%%%%%%%%%%%%%%%

{}From {}Fig.~\ref{fig3}, we see that $|\psi_k(t_{\rm max)}|$ can grow
to larger values the smaller $m_\chi$ is compared to $m_\phi$.  For
$m_\chi \ll m_\phi$, when the results approach a linear behavior shown
in {}Fig.~\ref{fig3a}, we have that $|\psi_k(t_{\rm max)}|$ can be
well approximated by
\begin{equation}
|\psi_k(t_{\rm max)}| \simeq \frac{m_\phi}{4\pi m_\chi \left(k^2
  +m_\chi^2\right)^{1/4}}.
\label{psikmax}
\end{equation} 
The results show that the condition $m_{\chi} \ll m_{\phi}$ is
required in order to have an efficient resonance.

{}For illustrative purpose, we also show in {}Fig.~\ref{fig3b} the
amplitude of $|\psi_k(t_{\rm max)}|$, when normalized by its value at
$k=0$, as a function of $k$.  As expected, we see that for $k <
m_{\chi}$ the growth rate is independent of $k$.

%%%%%%%%%%%%%%%%%FIGURE04%%%%%%%%%%%%%%%%%%%
\begin{center}
\begin{figure}[!htb]
\includegraphics[width=6.4cm]{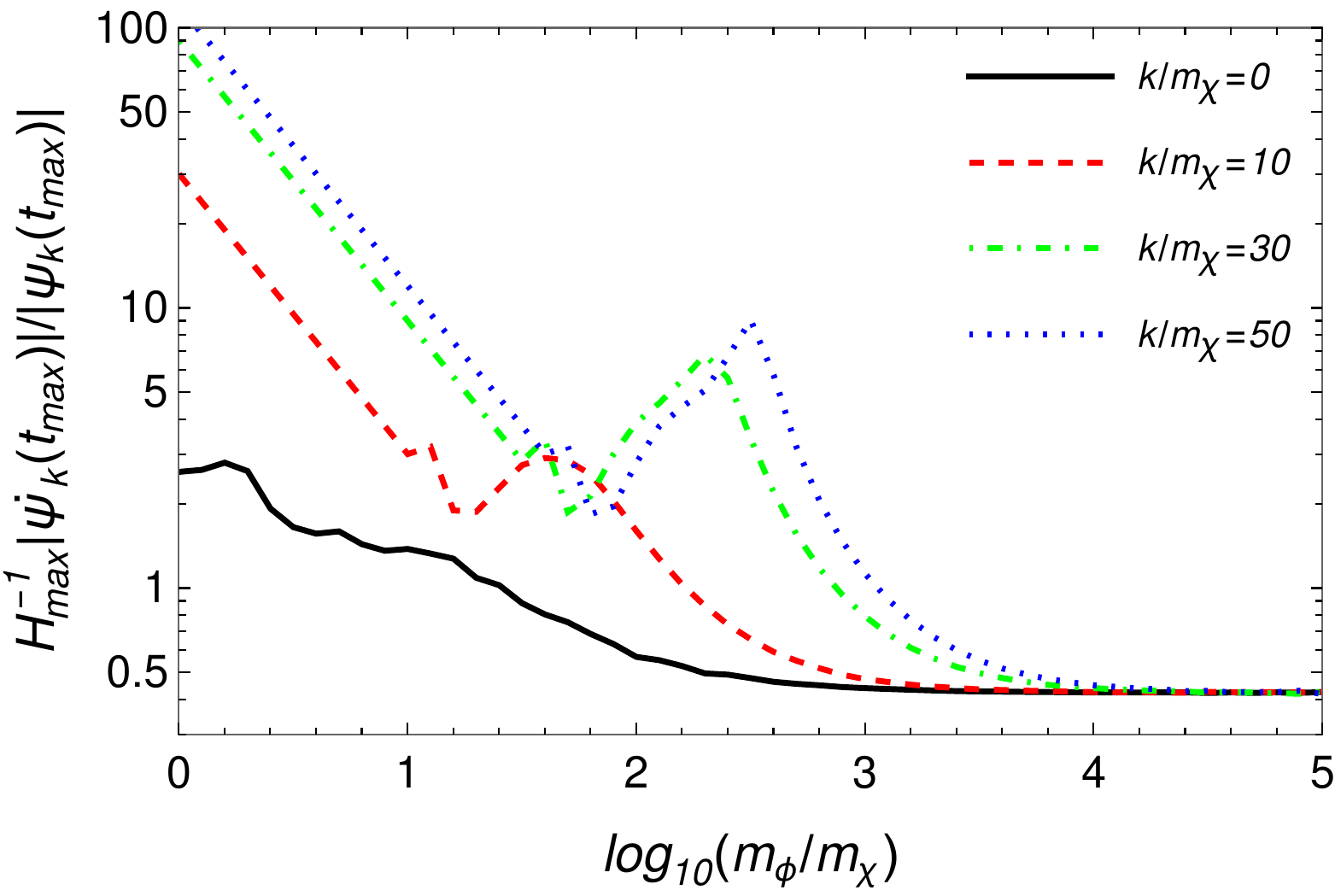}
\caption{The growth rate for the modes computed at $t=t_{\rm max}$
  normalized by the Hubble rate.}
  \label{fig4}
\end{figure}
\end{center}
%%%%%%%%%%%%%%%%%%%%%%%%%%%%%%%%%%%%%%%%%%%%

The rate of growth for the modes can be estimated by computing
\be 
\Gamma \, \equiv \, |\dot{\psi}_k/\psi_k| \, .  
\ee  
This rate is shown and compared to $H^{-1}$ in {}Fig.~\ref{fig4}.
Note that $\Gamma\gg H$, for values of $m_\chi$ which are not much
smaller than $m_\phi$. But this is a region where the modes are not
sufficiently excited.  In the region where the modes experience
pronounced growth (see {}Fig.~\ref{fig3}), we have $\Gamma \simeq
H/2$.  Thus, as to be expected from the analytical analysis of the
previous section, the resonance process is not fast on the Hubble time
scale. Note,  however, that the process is far more efficient than the
usual graviton mediated decay channel for which the rate is given by
the Eq.~(\ref{width}) shown in the following section.

%%%%%%%%%%%%%%%%%FIGURE05%%%%%%%%%%%%%%%%%%%
\begin{center}
\begin{figure}[!htb]
\subfigure[]{\label{fig5a} \includegraphics[width=6.4cm]{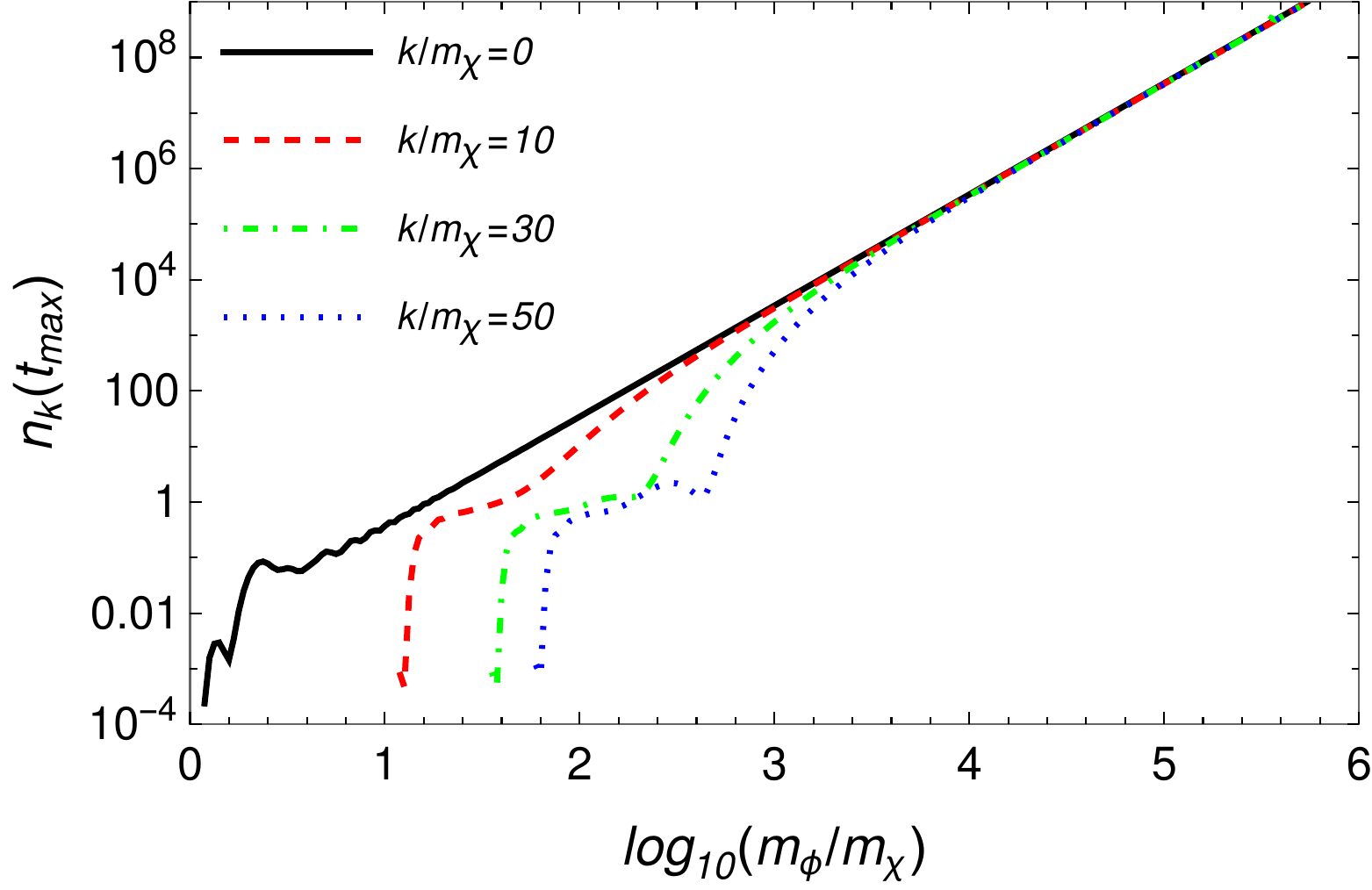}}
\subfigure[]{\label{fig5b} \includegraphics[width=6.4cm]{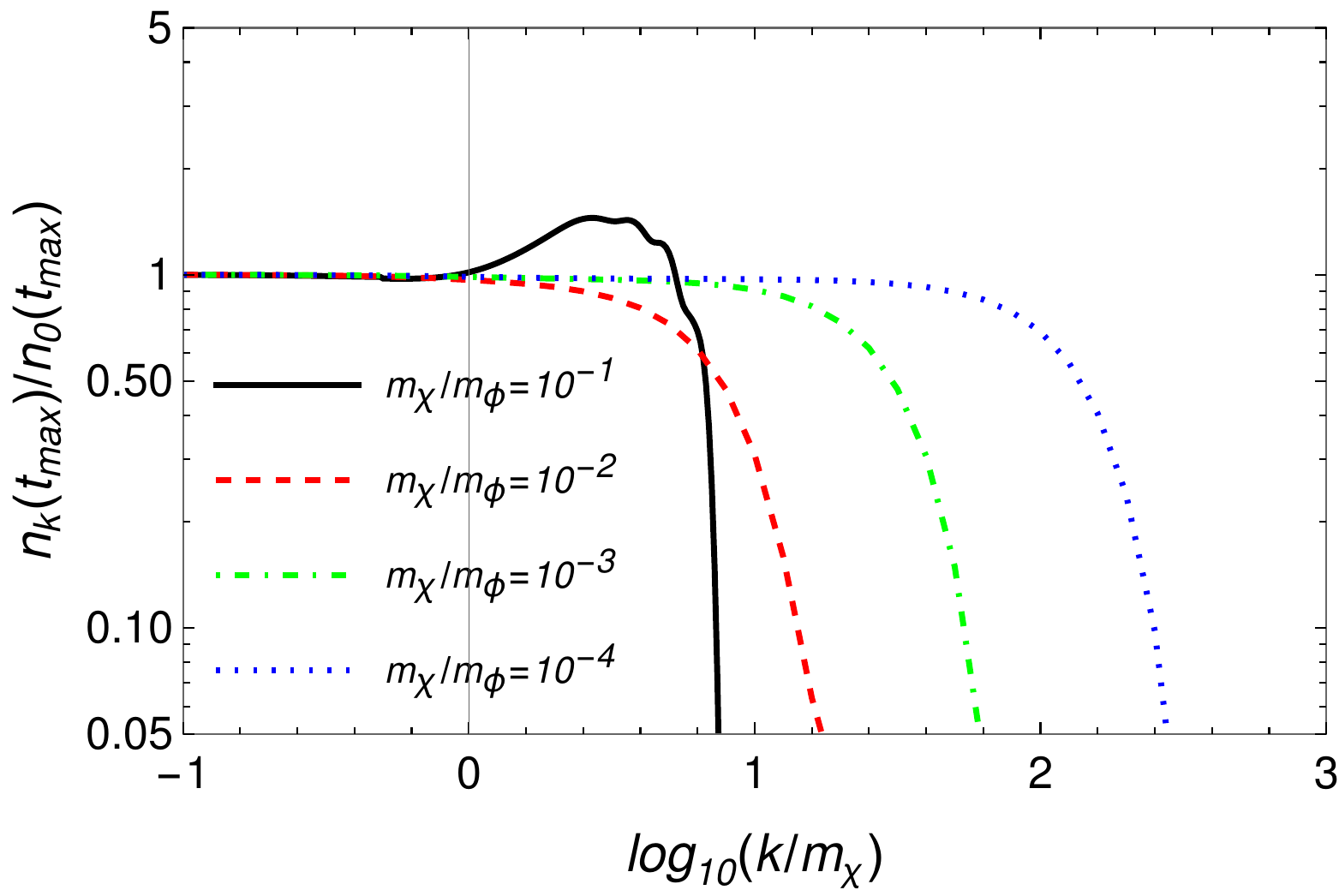}}
\caption{The comoving occupation number $n_k$ evaluated at $t=t_{\rm
    max}$.  $n_0$ denotes the occupation number evaluated at vanishing
  momentum.}
  \label{fig5}
\end{figure}
\end{center}
%%%%%%%%%%%%%%%%%%%%%%%%%%%%%%%%%%%%%%%%%%%%

To illustrate the relative growth of the $\chi$-momentum modes, we can
continue to work directly with the rescaled modes $\psi_k$, from which
we have the adiabatic invariant for Eq.~(\ref{psieq}) with the
interpretation of a comoving occupation number~\cite{KLS2},
\begin{equation}
n_k \,= \, \frac{\omega_k}{2} \left( \frac{|\dot
  \psi_k|^2}{\omega_k^2} + |\psi_k|^2\right) -\frac{1}{2}.
\label{nk}
\end{equation}
In F{}ig.~\ref{fig5}, we show the occupation number $n_k$ when
computed at $t=t_{\rm max}$.  As expected from the analytical
analysis, the larger the ratio of $m_{\phi} / m_{\chi}$, the longer
the period of resonance and the larger the value of $n_k$. Note that
for $m_\chi \ll m_\phi$ the resonance is shut off for values of  $k
\gg m_\chi$ such $m_\phi \gg k \gg m_\chi$.  The linear asymptotic
behavior seen in Fig.~\ref{fig5a} is well described by 
\be 
n_k(t_{\rm max})  \, \sim \,  \sqrt{k^2 +m_\chi^2} |\psi_k(t_{\rm
  max)}|^2/2 \, , 
\ee
with $|\psi_k(t_{\rm max)}|$ given by Eq.~(\ref{psikmax}).  As can
also be seen in {}Fig.~\ref{fig5a} and \ref{fig5b},  for $k$ modes in
the deep ultraviolet, the occupation number sharply drops as expected.

%%%%%%%%%%%%%%%%%%%%%%%%%%%%%%%%%%%%%%%%%%%%%%%%
\section{Comparison with Perturbative Effects}

In this section we wish to compare the resonant growth of $\chi$
particles with the perturbative growth which is obtained by the decay
of the $\phi$ condensate mediated by gravitons. This decay width for
this process was computed in~\cite{Mambrini:2021zpp} and takes the
form
\begin{equation}
\Gamma_{\phi\phi\to\chi\chi} = \frac{\rho_\phi m_\phi}{1024 \pi M_{\rm
    Pl}^4} \left( 1 + \frac{m_\chi^2}{2 m_\phi^2} \right)^2
\sqrt{1-\frac{m_\chi^2}{ m_\phi^2}},
\label{width}
\end{equation}
where $M_{\rm Pl}=1/\sqrt{8 \pi G}$ is the reduced Planck mass, with
$1/\sqrt{G}\equiv m_{\rm Pl}$ and $\rho_\phi$ is the energy density of
the inflaton field.  At the end of inflation (i.e., when the equation
of state for the inflaton becomes larger than $-1/3$), we have that
$\dot \phi_{\rm end}^2 = V(\phi_{\rm end})$ and then
\begin{equation}
\rho_{\phi,{\rm end}} = \frac{3}{2} V(\phi_{\rm end}) \, .
\label{rhophi}
\end{equation}
Assuming a quadratic potential for the inflaton field as in
Eq.~(\ref{pot}), we have
\be 
\phi_{\rm end}^2 \, \simeq \, M_{\rm Pl}^2/(4\pi) \equiv 2 M_{\rm
  Pl}^2 
\ee
and hence
\be 
\rho_{\phi,{\rm end}} \, =\, 3 m_\phi^2 M_{\rm Pl}^2 
\ee
and $H_{\rm end} = m_\phi$.  Recalling that for a quadratic inflaton
potential normalized to get the right amplitude of curvature
fluctuations we have 
\be 
m_\phi \, \approx \,  10^{-6} M_{\rm Pl} \, , 
\ee
we find that 
\be 
\Gamma_{\phi\phi\to\chi\chi}/H \, \sim \, 10^{-9} \, .  
\ee

{}From this discussion we conclude that the minimal preheating process
discussed in this paper is many orders of magnitude more efficient
that perturbative decay mediated by gravitons \footnote{Note, however,
there are many attempts of constructing (very model dependent)
scenarios of dark matter production through these type of
gravitational processes (see
e.g. Refs.~\cite{Mambrini:2021zpp,Clery:2021bwz,Haque:2021mab,Barman:2022qgt}
for some recent examples) which are more efficient than the above
perturbative decay.  Likewise, there are also attempts of fully
describe reheating after inflation using such gravitational decay
channels for the inflaton, like, e.g., in
Ref.~\cite{Haque:2022kez}. Here, we do not consider such cases since
our proposal is quite different from what was studied in those
works.}.

%%%%%%%%%%%%%%%%%%%%%%%%%%%%%%%%%%%%%%%%%%%%%%%%
\section{Conclusions}

We have proposed a model-independent minimal preheating process which
directly transfers the energy of the oscillating inflaton condensate
to Standard Model particles, in particular the Higgs field.  The
effect is a consequence of a parametric resonance instability of
tachyonic nature in the mode equation for scalar matter fields. This
resonance is due to an oscillating contribution to the mass term of
the modes which results from the oscillating pressure of the inflaton
condensate. We have established a lower bound on the efficiency of the
energy transfer.  It leads to the decay of the condensate on a time
scale given by (\ref{endtime}), which is larger than the Hubble time
at the end of inflation by a factor of the order of $10^3 (m_{\rm Pl}
/ m_{\phi})^{3/2}$, where $m_{\phi}$ is the mass of the inflaton
candidate.  The effective temperature at the end of the dominance of
the inflaton condensate is given by (\ref{Result}), which is smaller
than the maximal temperature one could have expected by a factor of
the order of $10^{-3/2}(G m_{\phi}^2)^{3/4}$.

Our analysis is based on an approximate analytical analysis in which
we chose our approximations in order to under-estimate the strength of
the effect. We confirmed the results using a numerical study of the
solutions of the mode equation.

Previous studies of reheating focused on matter field excitations via
matter couplings in the matter Lagrangian and are hence very model
dependent. For large matter couplings, these non-gravitational
channels can be more efficient than the one considered here, However,
since we do not expect Standard Model particles to be directly coupled
to the inflaton field, the previous mechanism leave open the question
of energy transfer to Standard Model particles. An important aspect of
our mechanism is that Standard Model particles are produced directly
(and they are produced more efficiently than particles of a Beyond the
Standard Model (BSM) if the masses of the BSM particles are much
larger than those of Standard Model particles).

We have focused on the production of scalar field quanta. The same
mechanism applies to gauge particle generation.  A similar instability
should also arise in the equation of motion for Fermions, but we
expect that Pauli blocking will suppress the strength of Fermion
particle production.

Our mechanism may have implications for the Dark Matter coincidence
problem, namely the question of why the current energy densities of
dark and visible matters are comparable.  If Dark Matter is a very
weakly interacting low mass (compared to the inflaton mass) scalar
field, then during the phase of oscillations of the inflaton
condensate the same energy density of dark matter and Higgs particles
will be generated.  How these densities evolve after the period of
inflaton oscillations will be model dependent, and we will return to
the study of this question in a future publication.

%%%%%%%%%%%%%%%%%%%%%%%%%%%%%%%%%%%%%%%%%%%%%
\begin{acknowledgements}

RB wishes to thank the Pauli Center and the Institutes of Theoretical
Physics and of Particle- and Astrophysics of the ETH for
hospitality. The research of RB at McGill is supported in part by
funds from NSERC and from the Canada Research Chair program.
V.K. would like to acknowledge the McGill University Physics
Department  for hospitality and partial financial support.
R.O.R. would like to thank the hospitality of the Department of
Physics McGill University.  R.O.R. also acknowledges financial support
of the Coordena\c{c}\~ao de Aperfei\c{c}oamento de Pessoal de
N\'{\i}vel Superior (CAPES) - Finance Code 001 and by research grants
from Conselho Nacional de Desenvolvimento Cient\'{\i}fico e
Tecnol\'ogico (CNPq), Grant No. 307286/2021-5, and from Funda\c{c}\~ao
Carlos Chagas Filho de Amparo \`a Pesquisa do Estado do Rio de Janeiro
(FAPERJ), Grant No. E-26/201.150/2021. 

\end{acknowledgements}

%%%%%%%%%%%%%%%%%%%%%%%%%%%%%%%%%%%%%%%%%%%%%


\begin{thebibliography}{99}

\bibitem{Guth}
A.~H.~Guth,
 ``The Inflationary Universe: A Possible Solution to the Horizon and Flatness Problems,''
 Phys.\ Rev.\ D {\bf 23}, 347 (1981)
 [Adv.\ Ser.\ Astrophys.\ Cosmol.\  {\bf 3}, 139 (1987)].
 doi:10.1103/PhysRevD.23.347;\\
 %%CITATION = doi:10.1103/PhysRevD.23.347;%%
 R.~Brout, F.~Englert and E.~Gunzig,
 ``The Creation Of The Universe As A Quantum Phenomenon,''
 Annals Phys.\  {\bf 115}, 78 (1978);\\
 %%CITATION = APNYA,115,78;%%
 A.~A.~Starobinsky,
 ``A New Type Of Isotropic Cosmological Models Without Singularity,''
 Phys.\ Lett.\ B {\bf 91}, 99 (1980);\\
 %%CITATION = PHLTA,B91,99;%%
 K.~Sato,
 ``First Order Phase Transition Of A Vacuum And Expansion Of The Universe,''
 Mon.\ Not.\ Roy.\ Astron.\ Soc.\  {\bf 195}, 467 (1981);\\
 %%CITATION = MNRAA,195,467;%%
L.~Z.~Fang,
  ``Entropy Generation in the Early Universe by Dissipative Processes Near the Higgs' Phase Transitions,''
  Phys.\ Lett.\  {\bf 95B}, 154 (1980).
  doi:10.1016/0370-2693(80)90421-9.
  %%CITATION = doi:10.1016/0370-2693(80)90421-9;%%
  
\bibitem{pert}
  %\cite{Dolgov:1982th}
A.~D.~Dolgov and A.~D.~Linde,
``Baryon Asymmetry in Inflationary Universe,''
Phys. Lett. B \textbf{116}, 329 (1982)
doi:10.1016/0370-2693(82)90292-1;\\
%\cite{Abbott:1982hn}
L.~F.~Abbott, E.~Farhi and M.~B.~Wise,
``Particle Production in the New Inflationary Cosmology,''
Phys. Lett. B \textbf{117}, 29 (1982)
doi:10.1016/0370-2693(82)90867-X;\\
%\cite{Albrecht:1982mp}
A.~Albrecht, P.~J.~Steinhardt, M.~S.~Turner and F.~Wilczek,
``Reheating an Inflationary Universe,''
Phys. Rev. Lett. \textbf{48}, 1437 (1982)
doi:10.1103/PhysRevLett.48.1437

\bibitem{TB}
J.~H.~Traschen and R.~H.~Brandenberger,
  ``Particle Production During Out-of-equilibrium Phase Transitions,''
  Phys.\ Rev.\ D {\bf 42}, 2491 (1990).
  doi:10.1103/PhysRevD.42.2491
  %%CITATION = doi:10.1103/PhysRevD.42.2491;%%
  
\bibitem{DK}
A.~D.~Dolgov and D.~P.~Kirilova,
  ``On Particle Creation By A Time Dependent Scalar Field,''
  Sov.\ J.\ Nucl.\ Phys.\  {\bf 51}, 172 (1990)
  [Yad.\ Fiz.\  {\bf 51}, 273 (1990)].
  %%CITATION = SJNCA,51,172;%%
  
\bibitem{KLS1}
L.~Kofman, A.~D.~Linde and A.~A.~Starobinsky,
``Reheating after inflation,''
Phys. Rev. Lett. \textbf{73}, 3195-3198 (1994)
doi:10.1103/PhysRevLett.73.3195
[arXiv:hep-th/9405187 [hep-th]].

\bibitem{STB}
%\cite{Shtanov:1994ce}
Y.~Shtanov, J.~H.~Traschen and R.~H.~Brandenberger,
``Universe reheating after inflation,''
Phys. Rev. D \textbf{51}, 5438-5455 (1995)
doi:10.1103/PhysRevD.51.5438
[arXiv:hep-ph/9407247 [hep-ph]].

\bibitem{KLS2}
L.~Kofman, A.~D.~Linde and A.~A.~Starobinsky,
``Towards the theory of reheating after inflation,''
Phys. Rev. D \textbf{56}, 3258-3295 (1997)
doi:10.1103/PhysRevD.56.3258
[arXiv:hep-ph/9704452 [hep-ph]].

\bibitem{RHrevs}
R.~Allahverdi, R.~Brandenberger, F.~Y.~Cyr-Racine and A.~Mazumdar,
  ``Reheating in Inflationary Cosmology: Theory and Applications,''
  Ann.\ Rev.\ Nucl.\ Part.\ Sci.\  {\bf 60}, 27 (2010)
  doi:10.1146/annurev.nucl.012809.104511
  [arXiv:1001.2600 [hep-th]];\\
  %%CITATION = doi:10.1146/annurev.nucl.012809.104511;%%
M.~A.~Amin, M.~P.~Hertzberg, D.~I.~Kaiser and J.~Karouby,
  ``Nonperturbative Dynamics Of Reheating After Inflation: A Review,''
  Int.\ J.\ Mod.\ Phys.\ D {\bf 24}, 1530003 (2014)
  doi:10.1142/S0218271815300037
  [arXiv:1410.3808 [hep-ph]].
  %%CITATION = doi:10.1142/S0218271815300037;%%
 
\bibitem{various}
%\cite{Ema:2015dka}
Y.~Ema, R.~Jinno, K.~Mukaida and K.~Nakayama,
``Gravitational Effects on Inflaton Decay,''
JCAP \textbf{05}, 038 (2015)
doi:10.1088/1475-7516/2015/05/038
[arXiv:1502.02475 [hep-ph]];\\
%\cite{Ema:2015oaa}
Y.~Ema, R.~Jinno, K.~Mukaida and K.~Nakayama,
``Particle Production after Inflation with Non-minimal Derivative Coupling to Gravity,''
JCAP \textbf{10}, 020 (2015)
doi:10.1088/1475-7516/2015/10/020
[arXiv:1504.07119 [gr-qc]];\\
 %\cite{Herranen:2015ima}
M.~Herranen, T.~Markkanen, S.~Nurmi and A.~Rajantie,
``Spacetime curvature and Higgs stability after inflation,''
Phys. Rev. Lett. \textbf{115}, 241301 (2015)
doi:10.1103/PhysRevLett.115.241301
[arXiv:1506.04065 [hep-ph]];\\
%\cite{Markkanen:2015xuw}
T.~Markkanen and S.~Nurmi,
``Dark matter from gravitational particle production at reheating,''
JCAP \textbf{02}, 008 (2017)
doi:10.1088/1475-7516/2017/02/008
[arXiv:1512.07288 [astro-ph.CO]];\\
%\cite{Ema:2016hlw}
Y.~Ema, R.~Jinno, K.~Mukaida and K.~Nakayama,
``Gravitational particle production in oscillating backgrounds and its cosmological implications,''
Phys. Rev. D \textbf{94}, no.6, 063517 (2016)
doi:10.1103/PhysRevD.94.063517
[arXiv:1604.08898 [hep-ph]];\\
%\cite{Ema:2018ucl}
Y.~Ema, K.~Nakayama and Y.~Tang,
``Production of Purely Gravitational Dark Matter,''
JHEP \textbf{09}, 135 (2018)
doi:10.1007/JHEP09(2018)135
[arXiv:1804.07471 [hep-ph]];\\
%\cite{Fairbairn:2018bsw}
M.~Fairbairn, K.~Kainulainen, T.~Markkanen and S.~Nurmi,
``Despicable Dark Relics: generated by gravity with unconstrained masses,''
JCAP \textbf{04}, 005 (2019)
doi:10.1088/1475-7516/2019/04/005
[arXiv:1808.08236 [astro-ph.CO]];\\
%\cite{Chung:2018ayg}
D.~J.~H.~Chung, E.~W.~Kolb and A.~J.~Long,
``Gravitational production of super-Hubble-mass particles: an analytic approach,''
JHEP \textbf{01}, 189 (2019)
doi:10.1007/JHEP01(2019)189
[arXiv:1812.00211 [hep-ph]];\\
%\cite{Ema:2019yrd}
Y.~Ema, K.~Nakayama and Y.~Tang,
``Production of purely gravitational dark matter: the case of fermion and vector boson,''
JHEP \textbf{07}, 060 (2019)
doi:10.1007/JHEP07(2019)060
[arXiv:1903.10973 [hep-ph]];\\
%\cite{Cembranos:2019qlm}
J.~A.~R.~Cembranos, L.~J.~Garay and J.~M.~S\'anchez Vel\'azquez,
``Gravitational production of scalar dark matter,''
JHEP \textbf{06}, 084 (2020)
doi:10.1007/JHEP06(2020)084
[arXiv:1910.13937 [hep-ph]];\\
%\cite{Babichev:2020yeo}
E.~Babichev, D.~Gorbunov, S.~Ramazanov and L.~Reverberi,
``Gravitational reheating and superheavy Dark Matter creation after inflation with non-minimal coupling,''
JCAP \textbf{09}, 059 (2020)
doi:10.1088/1475-7516/2020/09/059
[arXiv:2006.02225 [hep-ph]];\\
%\cite{Kaneta:2022gug}
K.~Kaneta, S.~M.~Lee and K.~y.~Oda,
``Boltzmann or Bogoliubov? Approaches compared in gravitational particle production,''
JCAP \textbf{09}, 018 (2022)
doi:10.1088/1475-7516/2022/09/018
[arXiv:2206.10929 [astro-ph.CO]].

%\cite{Alsarraj:2021yve}
\bibitem{Mesbah}
M.~Alsarraj and R.~Brandenberger,
``Moduli and graviton production during moduli stabilization,''
JCAP \textbf{09}, 008 (2021)
doi:10.1088/1475-7516/2021/09/008
[arXiv:2103.07684 [hep-th]].

\bibitem{we}
%\cite{Brandenberger:2023idg}
R.~Brandenberger, V.~Kamali and R.~O.~Ramos,
Decay of ALP Condensates via Gravitation-Induced Resonance,
[arXiv:2303.14800 [hep-ph]].

\bibitem{Natalia}
  N.~Shuhmaher and R.~Brandenberger,
  ``Non-perturbative instabilities as a solution of the cosmological moduli problem,''
  Phys.\ Rev.\ D {\bf 73}, 043519 (2006)
  doi:10.1103/PhysRevD.73.043519
  [hep-th/0507103].
  %%CITATION = doi:10.1103/PhysRevD.73.043519;%%  

\bibitem{Chunshan}
%\cite{Brandenberger:2022xbu}
R.~Brandenberger, P.~C.~M.~Delgado, A.~Ganz and C.~Lin,
``Graviton to photon conversion via parametric resonance,''
Phys. Dark Univ. \textbf{40}, 101202 (2023)
doi:10.1016/j.dark.2023.101202
[arXiv:2205.08767 [gr-qc]].

%\cite{Turner:1983he}
\bibitem{Turner:1983he}
M.~S.~Turner,
Coherent Scalar Field Oscillations in an Expanding Universe,
Phys. Rev. D \textbf{28}, 1243 (1983)
doi:10.1103/PhysRevD.28.1243
%784 citations counted in INSPIRE as of 20 Apr 2023

\bibitem{RHB-IC-Rev}
  %\cite{Brandenberger:2016uzh}
R.~Brandenberger,
``Initial conditions for inflation \textemdash{} A short review,''
Int. J. Mod. Phys. D \textbf{26}, no.01, 1740002 (2016)
doi:10.1142/S0218271817400028
[arXiv:1601.01918 [hep-th]].

\bibitem{Mathieu}
L. Landau and E. Lifshitz, {\it Mechanics} (3rd Edition) (Elsevier, 1982);\\
V. Arnold, {\it Mathematical Methods of Classical Mechanics} (2nd Edition) (Springer, New York, 1989);\\
N. W. McLachlan, {\it Theory and Applications of Mathieu Functions} (Oxford Univ. Press, Oxford, 1947).

%\cite{Mambrini:2021zpp}
\bibitem{Mambrini:2021zpp}
Y.~Mambrini and K.~A.~Olive,
Gravitational Production of Dark Matter during Reheating,
Phys. Rev. D \textbf{103}, no.11, 115009 (2021)
doi:10.1103/PhysRevD.103.115009
[arXiv:2102.06214 [hep-ph]].
%53 citations counted in INSPIRE as of 15 Mar 2023

%\cite{Clery:2021bwz}
\bibitem{Clery:2021bwz}
S.~Clery, Y.~Mambrini, K.~A.~Olive and S.~Verner,
Gravitational portals in the early Universe,
Phys. Rev. D \textbf{105}, no.7, 075005 (2022)
doi:10.1103/PhysRevD.105.075005
[arXiv:2112.15214 [hep-ph]].
%19 citations counted in INSPIRE as of 15 Mar 2023

%\cite{Haque:2021mab}
\bibitem{Haque:2021mab}
M.~R.~Haque and D.~Maity,
Gravitational dark matter: Free streaming and phase space distribution,
Phys. Rev. D \textbf{106}, no.2, 023506 (2022)
doi:10.1103/PhysRevD.106.023506
[arXiv:2112.14668 [hep-ph]].
%19 citations counted in INSPIRE as of 15 Mar 2023

%\cite{Barman:2022qgt}
\bibitem{Barman:2022qgt}
B.~Barman, S.~Cl\'ery, R.~T.~Co, Y.~Mambrini and K.~A.~Olive,
Gravity as a portal to reheating, leptogenesis and dark matter,
JHEP \textbf{12}, 072 (2022)
doi:10.1007/JHEP12(2022)072
[arXiv:2210.05716 [hep-ph]].
%2 citations counted in INSPIRE as of 09 Apr 2023

%\cite{Haque:2022kez}
\bibitem{Haque:2022kez}
M.~R.~Haque and D.~Maity,
Gravitational reheating,
Phys. Rev. D \textbf{107}, no.4, 043531 (2023)
doi:10.1103/PhysRevD.107.043531
[arXiv:2201.02348 [hep-ph]].
%17 citations counted in INSPIRE as of 09 Apr 2023

 

\end{thebibliography}
\end{document}